\documentclass[prb,twocolumn]{revtex4}
\usepackage{graphicx}

\begin{document}

\title{Pressure-induced metallization and structural phase transition of the Mott-Hubbard insulator TiOBr}

\author{C. A. Kuntscher,$^{1*}$ S. Frank,$^{1}$ A. Pashkin,$^{1}$ H. Hoffmann,$^{1}$ A. Sch\"onleber,$^{2}$ 
S. van Smaalen,$^{2}$ M. Hanfland,$^{3}$
S. Glawion,$^{4}$ M. Klemm,$^1$ M. Sing,$^4$ S. Horn,$^1$ and R. Claessen$^4$}
\address{
$^1$ Experimentalphysik 2, Universit\"at Augsburg, D-86135 Augsburg, Germany \\
$^2$ Laboratory of Crystallography, Universit\"at Bayreuth, 95440 Bayreuth, Germany \\
$^3$ European Synchrotron Radiation Facility, BP 220, F-38043 Grenoble, France \\
$^4$ Experimentelle Physik 4, Universit\"at W\"urzburg, D-97074 W\"urzburg, Germany}

\date{\today}

\begin{abstract}
We investigated the pressure-dependent optical response of the low-dimensional Mott-Hubbard 
insulator TiOBr by transmittance and reflectance measurements in the infrared and visible 
frequency range. A suppression of the transmittance above a critical pressure and a concomitant 
increase of the reflectance are observed, suggesting a pressure-induced metallization of TiOBr. 
The metallic phase of TiOBr at high pressure is confirmed by the presence of additional 
excitations extending down to the far-infrared range.
The pressure-induced metallization coincides with a structural phase transition, 
according to the results of x-ray powder diffraction experiments under pressure. 
\end{abstract}                    


\maketitle

\section{Introduction}
The titanium oxyhalides TiO$X$, with $X$=Cl or Br, are spin-Peierls compounds with exotic
properties.\cite{Seidel03,Kataev03,Ruckkamp05a,Krimmel05,Smaalen05,Shaz05,Fausti07}
The magnetic properties are related to the direct exchange interaction between
the spins on different Ti ions along the crystal axis $b$, forming an antiferromagnetic,
one-dimensional spin-1/2 Heisenberg chain.
With the electronic configuration $3d^1$ these materials are Mott-Hubbard insulators 
with a charge gap of $\approx$2~eV. \cite{Ruckkamp05a,Ruckkamp05b,Kuntscher06}
They have been earlier discussed to exhibit a resonating valence bond state and high-temperature 
superconductivity upon doping.\cite{Beynon93,Craco06}
However, up to now a metallization of TiO$X$ upon doping was not successful.\cite{Klemm08} 

Recent pressure-dependent infrared spectroscopic investigations on TiOCl suggest 
that the application of external pressure is an alternative way to induce an 
insulator-to-metal transition in TiO$X$: \cite{Kuntscher06}
Above 12~GPa the infrared transmittance is suppressed and the reflectance 
abruptly increases. Concomitantly, the sample color changes. 
From the pressure-dependence of the absorption edge the closure
of the charge gap at around 12~GPa was inferred. The pressure-induced changes
were attributed to the onset of additional excitations in the infrared frequency
range, suggesting an insulator-to-metal transition. 
However, the metallization of TiOCl could not be directly proved, since the spectroscopic
studies were restricted to the near-infrared range. Furthermore, due to the lack of 
pressure-dependent crystal structure data it could not be decided whether the 
presumable insulator-to-metal transition is bandwidth-controlled, i.e., whether it
is of pure electronic origin and can be described in the frame of the Mott 
transition,\cite{Imada98,Rozenberg96} or whether additionally changes in the 
crystal symmetry have to be considered.

To address these open issues, we have extended our pressure studies in several 
respects: (i) extension of the spectroscopic investigations to the far-infrared 
range, in order to verify the pressure-induced metallization; 
(ii) x-ray diffraction measurements under pressure, in order to clarify the origin of 
the observed pressure dependence of the optical response. Furthermore, we 
have carried out corresponding studies on the analog TiOBr. 

For TiOCl and TiOBr we found consistent spectroscopic results, namely evidence for the 
metallization of the materials under pressure. Furthermore, the metallization 
coincides with a structural phase transition induced by external pressure. 
In this Rapid Communication we focus on the material TiOBr. An extensive and 
comparative presentation of all our results for both compounds TiOBr and TiOCl 
will be presented in a forthcoming paper.\cite{Kuntscher08}

\section{Experimental}

Single crystals of Ti\nolinebreak OBr were synthesized by chemical vapor transport 
technique. TiOBr crystallizes in the space group $Pmmn$ consisting of buckled Ti-O 
bilayers parallel to the $ab$-plane and separated by layers of Br ions stacked along 
the $c$ direction.\cite{Schaefer58}

In our pressure-dependent studies diamond anvil cells (DACs) were used for the generation 
of pressures. The applied pressures were determined with the ruby luminescence method.\cite{Mao86} 
For the transmittance measurements argon was used as pressure medium. 
For the reflectance measurements finely ground CsI powder was chosen as pressure 
medium to insure direct contact of the sample with the diamond window.
For each transmittance and reflectance measurement a small piece (about 80 $\mu$$m$ $\times$
80 $\mu$$m$) was cut from single crystals with a thickness of $\leq$5 $\mu$$m$ and
placed in the hole of a steel gasket. 

Pressure-dependent transmittance and reflectance experi\-ments were
conducted at room temperature using a Bruker IFS 66v/S FT-IR
spectrometer with an infrared microscope (Bruker IRscope II). 
Part of the measurements were carried out at the infrared beamline of the synchrotron radiation source ANKA,
where the same equipment is installed. The reprodu\-cibility of the results was ensured 
by several experimental runs on different pieces of crystals. 
The geometries for the transmittance and reflectance measurements are the same
as used in Ref.~\onlinecite{Kuntscher06}.
The pressure-dependent transmittance was studied in the frequency range
2400 - 22000~cm$^{-1}$ (0.30 - 2.7~eV) for the polarization directions 
{\bf E}$||$$a,b$.
We measured the intensity I$_{\rm s}$($\omega$) of the radiation transmitted by the sample;
as reference, for each pressure we focused the incident radiation spot on the empty
space in the gasket hole next to the sample and obtained the transmitted intensity
I$_{\rm r}$($\omega$). The ratio $T$($\omega$)=I$_{\rm s}$($\omega$)/I$_{\rm r}$($\omega$)
is a measure of the transmittance of the sample.

Pressure-dependent reflectance measurements were carried out in the frequency range
$\approx$250 - 7500~cm$^{-1}$ (0.03 - 0.93~eV) for {\bf E}$||$$a,b$.
Reflectance spectra, $R_{\rm s-d}$, of the sample with respect to diamond were
obtained by measuring the intensity $I_{\rm s-dia}$($\omega$) reflected at the
interface between the sample and the diamond anvil.\cite{Kuntscher06}
As reference, the intensity $I_{\rm dia}$($\omega$) reflected from the inner diamond-air interface
of the empty DAC was used. The reflectance spectra were calculated according to
$R_{\rm s-d}(\omega)=R_{\rm dia}\cdot I_{\rm s-dia}(\omega)/I_{\rm dia}(\omega)$, where
$R_{\rm dia}$ was estimated from
the refractive index of diamond $n_{\rm dia}$ to 0.167 and assumed to be
independent of pressure.\cite{Eremets92,Ruoff94}

Pressure-dependent x-ray powder diffraction measurements at room temperature 
were carried out with monochromatic radiation ($\lambda$= 0.4128~\AA) at beamline
ID09A of the European Synchrotron Radiation Facility at Grenoble. Crystals were
ground and placed into a DAC for pressure generation. Helium served as hydrostatic
pressure-transmitting medium. Diffraction patterns were recorded with an 
image plate detector and then integrated\cite{Hammersley98} to yield an intensity vs 
2$\theta$ diagrams. The DAC was rotated by $\pm$3$^\circ$ during the exposure to 
improve the powder averaging. We carried out LeBail fits of the diffraction data using
the Jana2000 software,\cite{Petricek06} in order to determine the lattice parameters as a function 
of pressure. The analysis of the x-ray diffraction data is, however, complicated 
by the preferred orientation of the crystallites inside the DAC, with the $c$ crystal 
axis oriented perpendicular to the diamond anvil surface, i.e., along the direction of 
incidence of the x-radiation. Therefore, Rietveld refinements of the diffraction patterns
could not be carried out.

\begin{figure}[t]
\includegraphics[width=0.8\columnwidth]{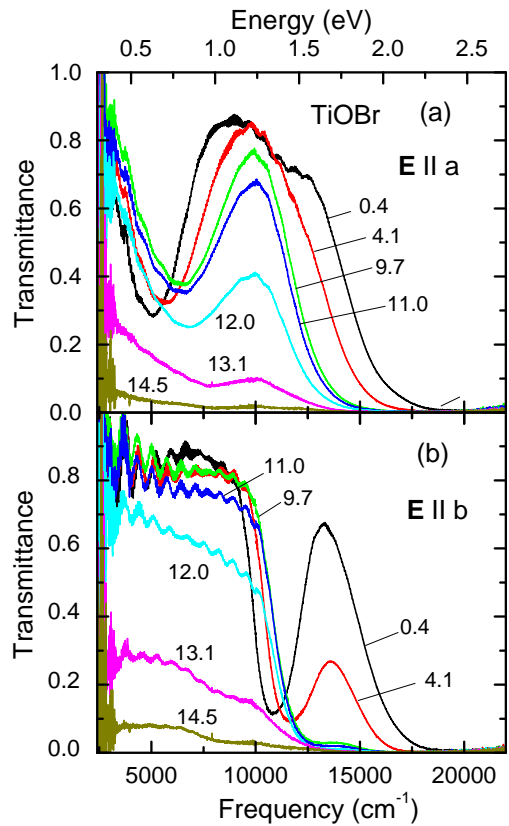}\\
\caption{(Color online) Room-temperature transmittance $T$($\omega$)=I$_{s}$($\omega$)/I$_{r}$($\omega$)
(see text for definitions) of TiOBr as a function of pressure for the polarization (a)
{\bf E}$||$$a$ and (b) {\bf E}$||$$b$ (pressure medium: argon). 
The numbers indicate the applied pressures in GPa.
} \label{fig:transmittance}
\end{figure}

\section{Results and Discussion}

In Fig.~\ref{fig:transmittance} we present the room-temperature transmittance spectra of 
TiOBr as a function of pressure for the polarizations {\bf E}$||$$a,b$.
The optical response of TiOBr resembles that of TiOCl:\cite{Kuntscher06}
For the lowest applied pressure the transmittance is suppressed above  $\approx$2~eV 
due to excitations across the charge gap. The absorption at 5060~cm$^{-1}$ (0.63~eV) for 
{\bf E}$||$$a$ and at 10950~cm$^{-1}$ (1.35~eV) for {\bf E}$||$$b$ are due to
excitations between the crystal field-split Ti $3d$ energy levels.\cite{Ruckkamp05a,Ruckkamp05b}
These excitations are infrared active due to the lack of inversion symmetry on the Ti sites;
they are slightly shifted in energy compared to TiOCl (for TiOCl the energies are 0.66
and 1.53~eV for {\bf E}$||$$a$ and {\bf E}$||$$b$, respectively).\cite{Kuntscher06}
The experimental values are in agreement with recent cluster calculations.\cite{Fausti07}

The pressure-induced changes (see Fig.~\ref{fig:transmittance}) are similar to those found 
for TiOCl:\cite{Kuntscher06}
With increasing pressure the orbital excitations broaden and continuously shift to higher 
frequencies. The absorption edge due to the excitations across the charge gap shifts 
to lower energies, and above $\approx$14~GPa the {\bf E}$||$$a,b$ transmittance is
suppressed over the whole studied frequency range. Like for TiOCl, a change of the 
sample color from red to black occurs upon pressure increase (not shown).
A quantitative presentation of the pressure dependence of the orbital excitations and
the charge gap in TiOBr will be given in a forthcoming paper.\cite{Kuntscher08}

\begin{figure}[t]
\includegraphics[width=0.85\columnwidth]{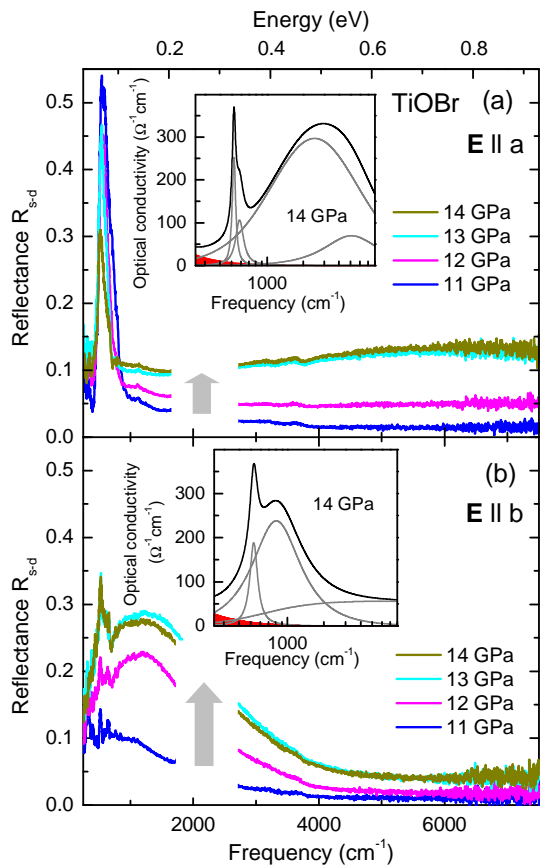}\\
\caption{(Color online) Room-temperature reflectance spectra $R_{\rm s-d}$($\omega$)
of TiOBr as a function of pressure for the polarization (a) {\bf E}$||$$a$ and
(b) {\bf E}$||$$b$ (pressure medium: CsI). Arrows indicate the changes with increasing pressure.
Insets: Real part of the optical conductivity spectra of TiOBr at the highest pressure (14~GPa)
obtained by fitting the reflectance spectra $R_{\rm s-d}$($\omega$) with the 
Drude-Lorentz model. Also shown are the different contributions (Drude term, Lorentz oscillators)
as obtained from the fit. The Drude contribution is indicated by the red (gray) area.}
 \label{fig:reflectance}
\end{figure}

Reflectance spectra measured at low pressures do not provide information about the reflectivity
of TiOBr, because at low pressures the thin TiOBr crystal in the DAC is partially transparent 
in the infrared range (see Fig.~\ref{fig:transmittance}), and interference fringes occur in
the reflectance spectra due to multiple reflections within the sample (data not shown).
On increasing pressure the interference fringes diminish and they have disappeared above 
$P$=10~GPa, at which pressure $R_{\rm s-d}$($\omega$) abruptly increases
in the whole studied frequency range (far-infrared to near-infrared), for both polarization directions.
The reflectance spectra $R_{\rm s-d}$($\omega$) in the high-pressure ($>$10~GPa) regime are
shown in Fig.~\ref{fig:reflectance}.
The difference in the critical pressure $P_{cr}$ derived from the transmittance ($P_{cr}=$13~GPa) 
and reflectance ($P_{cr}$=10~GPa) measurements are due to the different pressure media 
used.\cite{Frank06,Kuntscher06}
The observation of an enhanced reflectance of TiOBr in the infrared range above 10~GPa
is consistent with our earlier results for TiOCl.\cite{Kuntscher06} However, for TiOCl this
enhancement occurs at slightly higher (12~GPa) pressure,\cite{comment1} 
indicating that TiOBr is more sensitive to the application of pressure. This is also
suggested by the pronounced pressure-induced changes in the transmittance spectra of TiOBr 
(see Fig.~\ref{fig:transmittance}) occuring already in the low-pressure regime.

\begin{figure}[h]
\includegraphics[width=0.8\columnwidth]{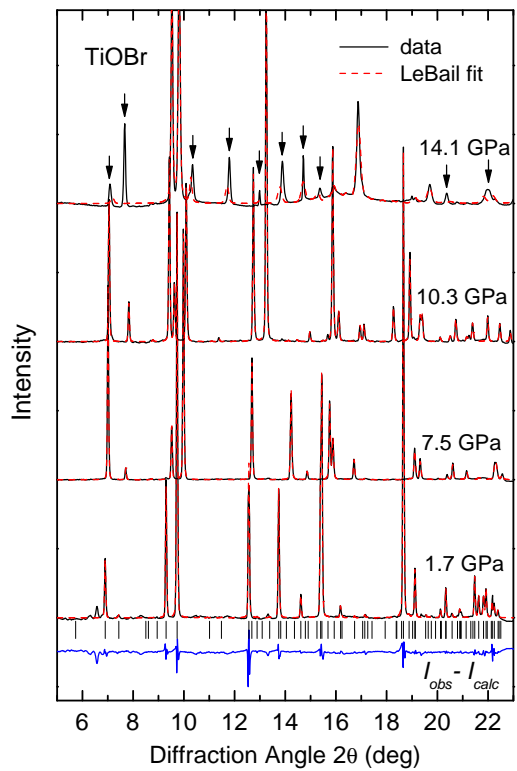}
\caption{(Color online) Room-temperature x-ray powder diffraction diagrams of TiOBr 
at high pressures ($\lambda$=0.4128~\AA) together with the LeBail fits (pressure medium: helium). 
For the lowest applied pressure (1.7~GPa) the difference curve ($I_{obs}-I_{calc}$) between 
the diffraction diagram and the LeBail fit is shown. 
Markers show the calculated peak positions for the ambient-pressure phase. Above 14~GPa the 
diffraction diagram can no longer be described by the ambient-pressure crystal symmetry.
Arrows indicate the diffraction peaks with the most obvious discrepancy between the 
data and the LeBail fitting curve.}
\label{fig:x-ray-TiOBr}
\end{figure}

Following the data analysis applied in Ref.~\onlinecite{Kuntscher06}, we fitted the reflectance spectra 
with the Drude-Lorentz model to obtain information about the pressure-induced excitations. The real
part of the optical conductivity of TiOBr at 14~GPa is presented in the insets of Fig.~\ref{fig:reflectance}
for the polarizations {\bf E}$||$$a,b$. At high pressure additional excitations occur, 
which extend down to the far-infrared range and include a Drude term (marked in red). 
This finding indicates the metallization of the sample under pressure.

Two scenarios were discussed\cite{Kuntscher06}  to explain the observed pressure-induced 
excitations inside the Mott-Hubbard gap of TiOCl:
a bandwidth-controlled insulator-to-metal transition, i.e., a Mott transition 
of purely electronic character,\cite{Rozenberg96,Imada98} 
and a structural phase transition entering a metallic phase. 
To clarify this issue, we carried out x-ray powder diffraction experiments
as a function of pressure. Fig.~\ref{fig:x-ray-TiOBr} shows the room-temperature diffraction 
diagrams of TiOBr for selected pressures together with the LeBail fits. 
Obviously, at low pressure the diffraction diagram can be well fitted with the 
ambient-pressure crystal structure (space group $Pmmn$). We could therefore obtain
the lattice parameters and unit cell volume as a function of pressure.\cite{Kuntscher08}
The most important finding, however, is that the diffraction diagram undergoes pronounced 
changes at around 14~GPa and is no longer compatible with the ambient-pressure crystal structure symmetry. 
These drastic changes in the diffraction data related to a structural phase transition
coincide with the changes in the optical properties described above. Corresponding results
are found for the analog compound TiOCl.\cite{Kuntscher08}
This points out that for a proper description of the pressure-induced transition in TiOBr and TiOCl
the lattice degree of freedom has to be taken into account besides correlation effects
in the electronic subsystem.

\section{Conclusion}

In conclusion, we studied the pressure dependence of the polarization-dependent 
optical response of the low-dimensional Mott-Hubbard insulater Ti\nolinebreak OBr
in the infrared and visible frequency range at room temperature.
The most interesting effects induced by pressure are the suppression
of the transmittance and the increase in the reflectance above a critical pressure,
suggesting the occurrence of an insulator-to-metal transition.
The metallization of TiOBr at high pressure is evidenced by the
finite optical conductivity in the far-infrared frequency range,
which includes a Drude term.
According to our pressure-dependent x-ray powder diffraction results,
the pressure-induced insulator-to-metal transition coincides 
with a structural phase transition for pressures around 14~GPa
under hydrostatic conditions. 
Our finding of a metallic phase in TiOBr under pressure raises the question whether
a superconduc\-ting state occurs in pressurized TiO$X$ ($X$=Br,Cl) at low temperature.

\subsection*{Acknowledgements}
Single-crystalline material of TiOBr was synthesized by A. Suttner (Bayreuth).
We acknowledge the ANKA Ang\-str\"om\-quelle Karlsruhe for the provision
of beamtime and we would like to thank B. Gasharova, Y.-L. Mathis,
D. Moss, and M. S\"upfle for assistance using the beamline ANKA-IR.
Financial support by the DFG, including the Emmy Noether-program, CL 124/6-1, 
and SFB 484, is gratefully acknowledged.


\begin{references}
\bibitem{comment1} 
This difference is related to the materials' properties, since the same pressure transmitting medium
(CsI) was used.
\end{references}
\end{document}